\documentclass[conference]{IEEEtran}
\IEEEoverridecommandlockouts
\usepackage{cite}
\usepackage{amsmath,amssymb,amsfonts,latexsym}
\usepackage{amsthm}
\usepackage{braket}
\usepackage{graphicx}
\usepackage{textcomp}
\usepackage{xcolor}
\usepackage{mathtools}
\usepackage{multirow}
\usepackage{url}
\usepackage{booktabs}
\usepackage{hyperref}
\usepackage{lipsum}
\usepackage{nicematrix}
\usepackage[all]{xy}
\usepackage{tikz-cd}
\usepackage{tikz}
\usepackage{makecell}
\usepackage{caption}
\usepackage{subcaption}
\usepackage{quantikz}

\usepackage{algorithm}
\usepackage{algpseudocode}

\usetikzlibrary{calc}
\usetikzlibrary{intersections,shapes.arrows}
\usetikzlibrary{arrows.meta, positioning, backgrounds}

\usepgfmodule{nonlineartransformations}

\newtheorem*{theorem*}{Theorem}

\theoremstyle{definition}
\newtheorem{example*}{Example}



\def\BibTeX{{\rm B\kern-.05em{\sc i\kern-.025em b}\kern-.08em
    T\kern-.1667em\lower.7ex\hbox{E}\kern-.125emX}}

\makeatletter
\newcommand{\linebreakand}{%
  \end{@IEEEauthorhalign}
  \hfill\mbox{}\par
  \mbox{}\hfill\begin{@IEEEauthorhalign}
}
\makeatother

\renewcommand{\figureautorefname}{Fig.~\negthinspace}
\renewcommand{\equationautorefname}{Eq.~\negthinspace}

\begin{document}

\title{Quantum Reinforcement Learning by Adaptive Non-local Observables\\

\thanks{
This work is supported by Laboratory Directed Research and Development Program \#24-061 of Brookhaven National Laboratory and National Quantum Information Science Research Centers, Co-design Center for Quantum Advantage (C2QA) under Contract No. DE-SC0012704. The research used resources of the NERSC, under Contract No. DE-AC02-05CH11231 using NERSC award HEPERCAP0033786. The views expressed in this article are those of the authors and do not represent the views of Wells Fargo. This article is for informational purposes only. Nothing contained in this article should be construed as investment advice. Wells Fargo makes no express or implied warranties and expressly disclaims all legal, tax, and accounting implications related to this article.}
}

\author{
\IEEEauthorblockN{Hsin-Yi~Lin}
\IEEEauthorblockA{\textit{Department of Mathematics} \\
\textit{and Computer Science} \\
\textit{Seton Hall University}\\
South Orange NJ, USA \\
hsinyi.lin@shu.edu}
\and
\IEEEauthorblockN{Samuel Yen-Chi~Chen}
\IEEEauthorblockA{\textit{Wells Fargo} \\
New York NY, USA \\
yen-chi.chen@wellsfargo.com}
\and
\IEEEauthorblockN{Huan-Hsin~Tseng}
\IEEEauthorblockA{\textit{AI \& ML Department} \\
\textit{Brookhaven National Laboratory}\\
Upton NY, USA  \\
htseng@bnl.gov}
\and 
\IEEEauthorblockN{Shinjae~Yoo}
\IEEEauthorblockA{\textit{AI \& ML Department} \\
\textit{Brookhaven National Laboratory}\\
Upton NY, USA \\
syjoo@bnl.gov}
}

\maketitle

\begin{abstract}
Hybrid quantum–classical frameworks leverage quantum computing for machine learning, but variational quantum circuits (VQCs) are limited by local measurements. We introduce an adaptive non-local observable (ANO) paradigm within VQCs for quantum reinforcement learning (QRL), jointly optimizing circuit parameters and multi-qubit measurements. The ANO-VQC architecture serves as the function approximator in Deep Q-Network (DQN) and Asynchronous Advantage Actor-Critic (A3C) algorithms. On multiple benchmark tasks, ANO-VQC agents outperform baseline VQCs. Ablation studies reveal that adaptive measurements enhance the function space without increasing circuit depth. Our results demonstrate that adaptive multi-qubit observables can enable practical quantum advantages in reinforcement learning.
\end{abstract}



\section{\label{sec:Indroduction}Introduction}

Quantum computing (QC) promises speed-ups for certain tasks that are intractable on classical hardware. Although current noisy intermediate-scale quantum (NISQ) devices suffer from decoherence, limited qubit counts, and gate infidelities, there has been remarkable progress in both hardware stabilization and error-mitigation techniques \cite{nielsen2010quantum, kandala2017hardware, preskill2018quantum}. These advances have motivated investigations into domains where quantum resources may offer concrete advantages over purely classical approaches.

One especially active direction is quantum machine learning (QML), which integrates quantum subroutines into classical learning pipelines to potentially enhance representation power, optimize high-dimensional landscapes, or accelerate kernel evaluations \cite{biamonte2017quantum, schuld2015introduction, havlivcek2019supervised}. Most QML schemes follow a hybrid quantum-classical paradigm, a parameterized quantum circuit (PQC) or variational quantum circuit (VQC) processes data, while a classical optimizer adjusts the circuit parameters to minimize a task-specific cost function \cite{mitarai2018quantum, benedetti2019parameterized,perez2020data,cerezo2021variational,bharti2022noisy}. This approach leverages quantum state preparation and entanglement, together with the maturity and flexibility of classical optimization loops. It was shown that VQC can represent complex distributions more efficiently than classical models \cite{du2020expressive, schuld2021effect}. Several works also observed that quantum learners require exponentially fewer queries and are more robust to noise, highlighting the potential of near-term quantum devices for machine learning tasks \cite{riste2017demonstration,fortunato2017noisy}.

Within quantum machine learning (QML), quantum reinforcement learning (QRL) explores the use of quantum circuits, particularly variational quantum circuits (VQCs), as policy or value-function approximators in reinforcement learning (RL) tasks \cite{meyer2022survey}. In RL, an agent interacts with an environment to maximize cumulative rewards, balancing exploration and exploitation. Classical algorithms such as Deep Q-Networks (DQN) \cite{mnih2013playing,mnih2015human} and Asynchronous Advantage Actor-Critic (A3C) \cite{mnih2016asynchronous} have been widely successful across a range of decision-making tasks.

Early QRL studies have demonstrated the feasibility of VQC-based agents in discrete control environments \cite{chen2020VQDQN,lockwood2020reinforcement,jerbi2021parametrized,chen2022variational, skolik2022quantum,chen2023asynchronous}. Hybrid quantum-classical agents using VQCs within a DQN framework have shown effective Q-value learning with fewer parameters than classical networks on tasks such as FrozenLake and cognitive radio control \cite{chen2020VQDQN}. VQCs integrated into DQN and Double DQN frameworks also achieved competitive or superior performance on environments like CartPole and Blackjack, supported by efficient quantum data encoding \cite{lockwood2020reinforcement}.

A variational quantum algorithm tailored for DQN was proposed in \cite{skolik2022quantum}, demonstrating that architectural choices such as input encoding and observable design significantly impact performance across discrete and continuous settings. In \cite{jerbi2021parametrized}, parameterized quantum policies were introduced, along with theoretical and empirical evidence indicating potential quantum advantage in specially constructed RL environments.

VQC-enhanced A3C was explored in \cite{chen2023asynchronous,kolle2024quantum}, where quantum agents achieved comparable or improved performance relative to classical baselines, with signs of faster convergence enabled by the expressiveness of quantum circuits and parallel learning strategies.

Conventionally, QNN architectures apply only local measurements, typically single-qubit Pauli operators, after the variational layers, which could restrict the accessible function space of the VQC and may hinder the learning of complex state–action correlations. To address this, researchers have introduced different measurement strategies, such as randomized measurements \cite{haug2023quantum, malmi2024enhanced} and pooling measurements in quantum convolutional neural networks (QCNNs) \cite{cong2019quantum}. More recently, a learnable observable framework was proposed to jointly optimize both circuit parameters and measurement bases \cite{chen2025learning,chen2025_Learning_To_Program_Measurement}, and this idea was further expanded into an adaptive non-local observable (ANO) paradigm \cite{lin2025adaptive} that dynamically selects entangled measurement operators to maximize model expressivity.

In this paper, we integrate the ANO models with VQC-based agents for RL frameworks, including DQN and A3C. The aim is to demonstrate that richer measurement schemes can significantly improve policy learning across diverse environments. Specifically, our contributions are as follows:
\begin{itemize}
    \item We incorporate adaptive non-local observables into the VQC backbone of both DQN and A3C agents, enabling the measurement scheme to be trained simultaneously with circuit parameters.
    \item  We evaluate our ANO-VQC agents on multiple controlled tasks (e.g., CartPole, MiniGrid, MountainCar), comparing performance to baseline VQC agents with local measurements.

    \item Through ablation studies, we show that ANO enhances the reachable function space of the VQC, leading to faster convergence and higher cumulative rewards.
\end{itemize}

Our results suggest that incorporating adaptive multi-qubit measurements can unlock latent potential in hybrid quantum-classical agents, thereby bringing us closer to realizing practical quantum advantages in reinforcement learning.

\section{Adaptive Non-local Observables for Quantum Reinforcement Learning}\label{sec_qrl}

\subsection{RL Formulation}\label{subsec_rl_formulation}

RL is a computational approach to learning through interaction with an environment, formulated by a Markov decision process (MDP) of a tuple $(S, A, P, \gamma, R)$, where $\mathcal{S}$ is the state space, representing all possible configurations of the environment; $\mathcal{A}$ is the action space, representing all possible actions the agent can take. $P: \mathcal{S} \times \mathcal{A} \times \mathcal{S} \rightarrow [0, 1]$ is the transition probability with $P(s'|s,a)$ denoting the transition probability from state $s \to s'$ under action $a$. Reward $R: \mathcal{S} \times \mathcal{A} \times \mathcal{S} \rightarrow \mathbb{R}$ is a function endowing value to the state transition $s \overset{a}{\to} s'$. Constant $\gamma \in [0, 1]$ is the discount factor determining the decaying rate of accumulated rewards.

A sequence of actions acting on an initial state $s_0 \in \mathcal{S}$ leads to the dynamics of an MDP:
\begin{equation}\label{E: s sequence}
    s_0 \xrightarrow{a_0} s_1 \xrightarrow{a_1} s_2 \xrightarrow{a_2} \cdots
\end{equation}
A \textit{policy} is a function $\pi: \mathcal{S} \rightarrow \mathcal{A}$ to generate action $\pi(s) \in \mathcal{A}$ under arbitrary $s \in \mathcal{S}$ given. In the context of RL, a policy $\pi$ is also called an \textit{agent}. The \textbf{goal} of RL is to find the \textit{optimal} policy $\pi^*$ deriving highest rewards along sequence \equationautorefname{\ref{E: s sequence}.

Given a policy $\pi$, one defines the \textit{state-value} function,
\begin{equation}\label{E: V}
    V^{\pi}(s) = \mathbb{E}_{\pi}\left[\sum_{t=0}^{\infty} \gamma^t R(s_t, a_t, s_{t+1}) \, | \, s_0 = s\right]
\end{equation}
to evaluate the advantage of state $s$. Another variant called $Q$-function is defined,
\begin{equation}\label{E: Q}
    Q^{\pi}(s, a) = \mathbb{E}_{\pi}\left[\sum_{t=0}^{\infty} \gamma^t R(s_t, a_t, s_{t+1}) \, | \, s_0 = s, a_0 = a\right]
\end{equation}
It is important to note that the above two functions depend on a policy $\pi$ given. The \textbf{$Q$-learning} reverses the process of \equationautorefname{\ref{E: Q}} to obtain the \textit{optimal} policy $\pi^*$ via
\begin{equation}\label{E: opt Q}
    \pi^*(s) = \arg\max_{a \in \mathcal{A}} Q^*(s, a)
\end{equation}
Consequently, $\pi^*$ is \textit{implicitly} derive via $Q^*$, while $Q^*$ is to be independently estimated by \textit{Bellman equation},
\begin{equation}\label{E: Bellman}
    Q^*(s,a) = R(s,a) +  \gamma \, \mathbb{E}_{s' \sim P(\cdot \, | \, s, a)} \left[ \max_{a' \in \mathcal{A}} Q^*(s', a') \right]
\end{equation}
Finding $Q^*$ to solve \equationautorefname{\ref{E: Bellman}} is therefore a critical task. The DQN~\cite{mnih2015human} is an approach using deep neural networks to approximate $Q^*$ through iterative optimization across numerous \textit{episodes}.

In contrast to the $Q$-learning, the A3C~\cite{mnih2016asynchronous} directly parameterize a policy with a value function,
\begin{equation}\label{E: actor-critic}
    \pi_{\theta}(a\mid s)\quad\text{(stochastic actor)},\quad V_{\psi}(s)\quad\text{(state–value critic)}
\end{equation}

with two sets of parameters $\theta \in \mathbb{R}^{d_\pi}$ and $\psi \in \mathbb{R}^{d_v}$. 

For a trajectory segment of at most $n$ steps collected by a worker thread, define the $n$-step return,
\[
G_t^{(n)} := \sum_{i=0}^{n-1}\gamma^{\,i} r_{t+i} + \gamma^{\,n}\,V_{\psi}(s_{t+n}), \quad
A_t = G_t^{(n)}-V_{\psi}(s_t).
\]
where $A_t$ is the \emph{advantage} estimate. Each worker minimizes the composite loss over its fragment,
\begin{equation}\label{E: A3C total loss}
    \mathcal{L} =\sum_{t=t_0}^{t_0 + n -1} \Bigl( \mathcal{L}_{\pi}(t) + c_v \, \mathcal{L}_{V}(t) + \mathcal{L}_{\mathrm{entropy}}(t) \Bigr)
\end{equation}
where $c_v>0$ weights the value loss and the losses at each step $t$ are given by,
\begin{equation}
    \begin{aligned}
\mathcal{L}_{\pi}(t)  &= -\log\pi_{\theta}(a_t\!\mid\!s_t)\, A_t,\\
\mathcal{L}_{V} (t)  &= \frac{1}{2} \bigl(G_t^{(n)}-V_{\psi}(s_t)\bigr)^2,\\
\mathcal{L}_{\mathrm{entropy}}(t) &= -\beta \, \sum_{a\in\mathcal{A}} \pi_{\theta}(a\!\mid\!s_t)\log\pi_{\theta}(a\!\mid\!s_t),
\end{aligned}
\end{equation}
where $\beta>0$ encourages exploration. Finally, each parallel worker asynchronously computes its local gradients $\nabla_{\theta}\mathcal{L}$, $\nabla_{\psi}\mathcal{L}$ to the \emph{shared} parameters $(\theta,\psi)$ with step‐size $\alpha>0$, then refreshes its local copy from the global variables. 

The combination of \textit{on-policy} (trajectory \equationautorefname{\ref{E: s sequence}} sampling from $\pi$) advantage gradients, entropy regularization, and lock-free asynchronous updates across workers constitutes the core of the A3C.

In this work, we build on the paradigms of $Q$-learning and A3C by modeling them with ANO to evaluate its reinforcement learning capabilities.

\subsection{Adaptive Non-local Observables for Quantum RL}\label{subsec_ano_qrl}

The variational quantum circuit (VQC), serving as a quantum neural network (QNN) in quantum machine learning (QML), is defined as
\begin{equation} \label{E: VQC}
    f_{\text{VQC}}(x) = \bra{\psi_0} W^{\dagger}(x)\, U^{\dagger}(\theta)\, H\, U(\theta)\, W(x) \ket{\psi_0},
\end{equation}
where the input \( x \in \mathbb{R}^n \) is encoded by an encoding layer \( W(x) \) acting on the initial state \( \ket{\psi_0} = \ket{0}^{\otimes n} \), followed by a parameterized variational layer \( U(\theta) \), see \figureautorefname{\ref{fig: ANO + VQC}}.

The encoding layer \( W(x) \) applies Hadamard gates to each qubit, followed by single-qubit rotations \( R(x_i) \), where the angles are determined by the input components \( x_i \). The variational layer \( U(\theta) \) consists of entangling CNOT gates between neighboring qubits and local rotations \( R(\theta_i) \) with learnable parameters \( \theta_i \). Finally, the observable \( H \) is measured to obtain the circuit output.

One limiting factor of VQC in \equationautorefname{\ref{E: VQC}} is due to the fixed Hermitian $H$ to result in a confined output, $\lambda_1 \leq f_{\text{VQC}}(x) \leq \lambda_n$,
where $\lambda_1 \leq \cdots \leq \lambda_n$ are the eigenvalues of $H$. 

VQCs using conventional \textit{Pauli observables} have only two eigenvalues $\lambda = \pm 1$, which further restricts the range as, 
\begin{equation}
    -1 \leq f_{\text{VQC}}(x) \leq 1
\end{equation}


To overcome the limitations imposed by fixed observables, particularly the constrained output range and limited expressivity of Pauli-based measurements, we propose integrating the ANO~\cite{lin2025adaptive} into the Quantum RL framework. ANO enhances VQC-based models by incorporating an adaptive $k$-local observable to reach more flexible and expressive function approximation.

In ANO, a $k$-local observable (with $k \leq n$ qubits) is parametrized by,
\begin{equation}\label{ano}
  H(\phi) =   \begin{pmatrix}
c_{11} & a_{12} + i b_{12} & a_{13} + i b_{13} & \cdots & a_{1K} + i b_{1K}  \\
* & c_{22}  & a_{23} + i b_{23}  & \cdots & a_{2K} + i b_{2K}  \\
* & * & c_{33}  & \cdots & a_{3K} + i b_{3K}  \\
\vdots & \vdots & \vdots & \ddots & \vdots \\
* & * & * & \cdots & c_{KK}
\end{pmatrix}
\end{equation}
with $K = 2^k$ and $\phi = \left( a_{ij}, b_{ij}, c_{ii} \right)_{i, j=1}^K$ denoting trainable parameters. The lower triangle entries are defined by complex conjugates to ensure $H(\phi) = H^{\dagger}(\phi)$.

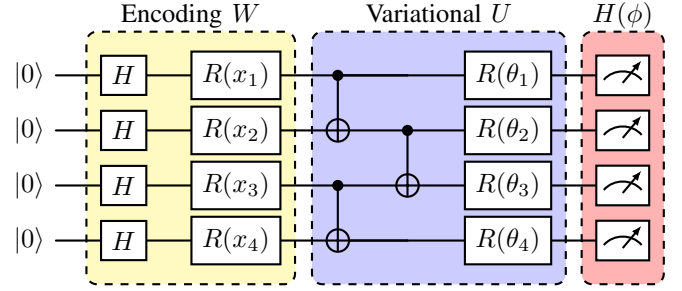
\begin{figure}[htbp]
 \vskip -0.1in
  \centering
    \begin{quantikz}[row sep=0.1cm, column sep=0.6cm]
        \lstick{$\ket{0}$} &\gate{H} \gategroup[wires=4,steps=2,style={dashed,rounded corners,fill=yellow!30,inner xsep=2pt},background]{Encoding $W$} & \gate{R(x_1)} & \ctrl{1} \gategroup[wires=4,steps=3,style={dashed,rounded corners,fill=blue!20,inner xsep=2pt},background]{Variational $U$} & \qw &\gate{R(\theta_1)}  & \meter{}\gategroup[4,steps=1,style={dashed,rounded corners,fill=red!30,inner xsep=2pt},background]{$H(\phi)$} \\
        \lstick{$\ket{0}$} &\gate{H} & \gate{R(x_2)} & \targ{}  & \ctrl{1} &\gate{R(\theta_2)}  & \meter{} \\
        \lstick{$\ket{0}$} &\gate{H} & \gate{R(x_3)} & \ctrl{1} & \targ{}  &\gate{R(\theta_3)}  & \meter{} \\
        \lstick{$\ket{0}$} &\gate{H}  & \gate{R(x_4)} & \targ{}  & \qw      &\gate{R(\theta_4)}  & \meter{}
    \end{quantikz}
  \caption{Illustration of ANO with VQC, described in \equationautorefname{\ref{E: VQC}} and \equationautorefname{\ref{ano}}. This is an example of a 4-qubit system.}
  \label{fig: ANO + VQC}
\end{figure}

For DQN, the ANO model approximates the action-value function \( Q(s,a) \) by evaluating
\begin{equation} \label{E: ANO function}
    f_{\theta,\phi}(s) = \bra{\psi_0} W^{\dagger}(s)\, U^{\dagger}(\theta)\, H(\phi)\, U(\theta)\, W(s) \ket{\psi_0},
\end{equation}
where the input state \( s \) is encoded via a unitary \( W(s) \), followed by a variational circuit \( U(\theta) \), and measured through an adaptive observable \( H(\phi) \). The model outputs a vector over all possible actions,  $f_{\theta,\phi}(s) = (Q(s, a_1), \ldots, Q(s, a_k)) \in \mathbb{R}^{|\mathcal{A}|} \to (Q^*(s,a_1), \ldots, Q^*(s,a_k))$, where $Q^*$ is the optimal action-value function for any $s$ and $a$.

The parameters \( \theta \) (for variational unitary rotations) and \( \phi \) (for observable design) are optimized by minimizing the Bellman loss:
\begin{multline}
L(\theta, \phi) = \mathbb{E}_{(s,a,s')\sim \mathcal{D}} \left[ \left( R(s, a) + \gamma \max_{a'} Q_{\theta',\phi'}(s',a') \right. \right. \\
\left. \left. - Q_{\theta,\phi}(s,a) \right)^2 \right],
\end{multline}
where \( \mathcal{D} \) is a replay buffer of transitions \( (s, a, s') \sim P(s' \mid s, a) \), and \( (\theta', \phi') \) are parameters of the target network.

Analogously, the A3C actor-critic pair in \equationautorefname{\ref{E: actor-critic}} are estimated by \equationautorefname{\ref{E: ANO function}} with Gibbs transformation,
\begin{equation}\label{E: ANO-A3C1}
    \pi_{\theta,\phi}(a | s) = \frac{\exp \left( f_{\theta,\phi}(s,a) \right)}{\sum_{a'\in\mathcal A} \exp \left( f_{\theta,\phi}(s,a') \right)}
\end{equation}
and the state–value critic is given by,
\begin{equation}\label{E: ANO-A3C2}
    V_{\vartheta, \varphi}(s)= \bra{\psi_0} W^{\dagger}(s)\,U^{\dagger}(\vartheta)\,H(\varphi)\,U(\vartheta)\,W(s) \ket{\psi_0}.
\end{equation}
with other parameters $\vartheta$ in the variational layer and $\varphi$ parameterizing another observable $H(\varphi)$, so that the parameters of the ANO model $(\theta, \vartheta,\phi,\varphi)$ minimize the loss \equationautorefname{\ref{E: A3C total loss}}.

Our motivation is to apply the strengths of ANO to RL for the following reasons.
\begin{enumerate}
    \item 
    ANO yields a wider range of value outputs. By the Rayleigh Quotient,
    \[
    \lambda_1 \leq \bra{\psi} H(\phi) \ket{\psi} \leq \lambda_n,
    \]
    where $\lambda_1 \leq \cdots \leq \lambda_n$ are \textit{flexible} spectrum of adaptable $H(\phi)$, which is useful in RL where $Q$-values can span arbitrary scales.

    \item
    A $k$-local ANO with limited connectivity ($k < n$) can express more complex measurement operators than fixed Pauli bases, allowing more accurate functional approximations.

    \item
    ANO adapts its representational capacity by jointly optimizing the variational parameters $\theta$ and observable parameters $\phi$, adjusting naturally to the complexity of the task.
\end{enumerate}

\section{Experiments}\label{sec_exp_results}

\subsection{ANO-DQN}

To evaluate the proposed ANO-enhanced Q-learning framework, we conduct experiments on the CartPole and Mountain Car environments using a VQC setup as defined in \equationautorefname{\ref{E: ANO function}}.

\subsubsection{\textbf{Cart-Pole}}

Defined by a continuous state space $\mathcal{S} \subset \mathbb{R}^4$ with $s = (x, \dot{x}, \theta, \dot{\theta}) \in \mathcal{S} $ representing the cart position $x$, 
velocity $\dot{x}$, pole angle $\theta$, and angular velocity $\dot{\theta}$. The action space is discrete: $\mathcal{A} = \{0, 1\}$, corresponding to applying a force left ($a = 0$) or right ($a = 1$).  At each time step, the agent receives a reward $R(t) = 1$ as long as the pole remains upright and the cart stays within bounds; the episode terminates otherwise.

Our results are summarized in \figureautorefname{\ref{fig: dqn: cartpole}}, comparing three configurations: 
\begin{enumerate}
    \item \textbf{(3-local w/ R.)} 3-local ANO with variational layer, where both variational unitary $U(\theta)$ and adaptive observable $H(\phi)$ are trained to approximate the $Q$-function,
    \item \textbf{(Only R.)} Conventional VQCs with Pauli Z measurement with variational  $U(\theta)$ but using \textit{fixed Pauli} matrix,
    \item \textbf{(Only measurement}) 3-local ANO without rotation, which removes the variational $U(\theta)$ and relies \textit{solely} on the trainable $H(\phi)$ for learning expressivity. 
\end{enumerate}

The 3-local configuration with rotation gates learns fastest and reaches the environment’s 500-step cap earlier and more stably than the other two. The traditional VQC (``only R.'') converges more slowly and to a lower average reward, similar to the measurement-only variant.

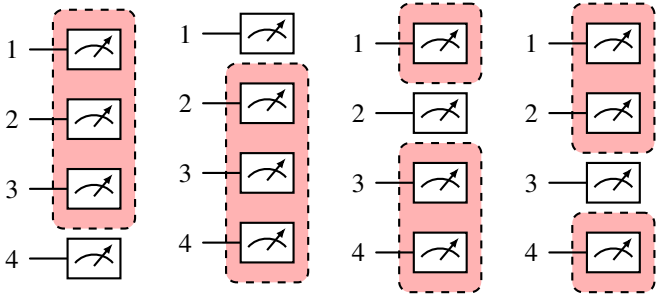
\begin{figure}[!h]
 \vskip -0.0in
  \centering
  \begin{tikzpicture}[scale=0.7, node distance=1.5cm]
    \node (circuit1) {
    \begin{quantikz}[row sep=0.4cm, column sep=0.5cm]
        \lstick{1}  &\meter{} \gategroup[3,steps=1,style={dashed,rounded corners,fill=red!30,inner xsep=2pt},background]{}\\
        \lstick{2} &\meter{} \\
        \lstick{3} &\meter{}\\
        \lstick{4} &\meter{}
    \end{quantikz}
    };
    \node (circuit2) [right=0.2cm of circuit1] {
    \begin{quantikz}[row sep=0.4cm, column sep=0.5cm]
        \lstick{1}  &\meter{} \\
        \lstick{2}  &\meter{} \gategroup[3,steps=1,style={dashed,rounded corners,fill=red!30,inner xsep=2pt},background]{}\\
        \lstick{3}  &\meter{}\\
        \lstick{4}  &\meter{}
    \end{quantikz}
    };
    \node (circuit3) [right=0.2cm of circuit2] {
    \begin{quantikz}[row sep=0.4cm, column sep=0.5cm]
        \lstick{1}  &\meter{} \gategroup[1,steps=1,style={dashed,rounded corners,fill=red!30,inner xsep=2pt},background]{}\\
        \lstick{2} &\meter{} \\
        \lstick{3}  &\meter{} \gategroup[2,steps=1,style={dashed,rounded corners,fill=red!30,inner xsep=2pt},background]{}\\
        \lstick{4} &\meter{}
    \end{quantikz}
    };
        \node (circuit4) [right=0.2cm of circuit3] {
    \begin{quantikz}[row sep=0.4cm, column sep=0.5cm]
        \lstick{1}  &\meter{} \gategroup[2,steps=1,style={dashed,rounded corners,fill=red!30,inner xsep=2pt},background]{}\\
        \lstick{2} &\meter{} \\
        \lstick{3}  &\meter{}\\
        \lstick{4} &\meter{} \gategroup[1,steps=1,style={dashed,rounded corners,fill=red!30,inner xsep=2pt},background]{}
    \end{quantikz}
    };
  \end{tikzpicture}
  \caption{Illustration of $3$-local ANO measurement on a $4$-qubit system. This example shows four distinct groupings of measured qubits. For an action space of dimension $a$, the first $a$ outputs from these groupings are used as the action logits.}
  \label{fig: sliding k-local}
\end{figure}

\begin{figure}[htbp]
\vskip -0.15in
\begin{center}
\centerline{\includegraphics[width=1.04\columnwidth]{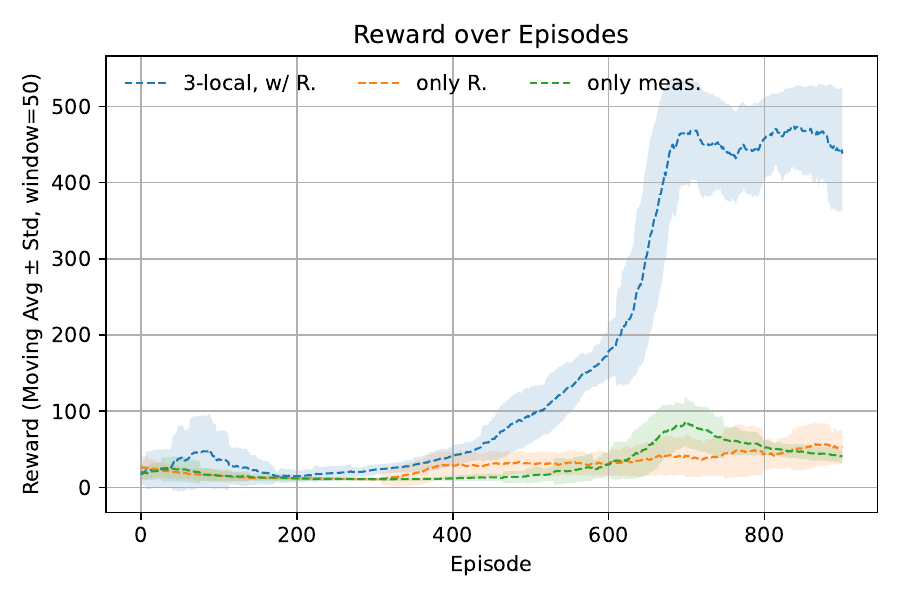}}
\caption{\textbf{[DQN] Cart-pole reward over episodes.} 
}
\label{fig: dqn: cartpole}
\end{center}
\vskip -0.2in
\end{figure}

\subsubsection{\textbf{Mountain Car}}

The Mountain Car environment is defined by a continuous state space $\mathcal{S} \subset \mathbb{R}^2$, where a state $s = (x, \dot{x}) \in \mathcal{S}$ represents the car's position $x \in [-1.2, 0.6]$ and velocity $\dot{x} \in [-0.07, 0.07]$. The action space is discrete $\mathcal{A} = \{0, 1, 2\}$, corresponding to left acceleration ($a=0$), no force ($a=1$), and right acceleration ($a=2$). The goal is to reach $x \geq 0.5$ by building momentum through repeated transitions $s_{t+1} \sim P(\cdot \mid s_t, a_t)$ under energy constraints.

To evaluate the influence of locality and variational rotation gates in the ANO model. To accommodate different locality levels, the original 2-dimensional input $s \in \mathbb{R}^2$ is duplicated once for the 3-local case (yielding 4 features encoded in 4 qubits), and twice for the 6-local and rotation-only settings (yielding 6 features encoded in 6 qubits). This setup allows for a controlled comparison of the model’s representational power under varying combinations of variational gates and measurement locality. 4 configurations using quantum approximator $ f_{\theta, \phi}(s, a) $ defined in \equationautorefname{\ref{E: ANO function}} are compared:
\begin{enumerate}
    \item \textbf{(3-local w/ R.)} uses a $k = 3$ adaptive observable $ H(\phi) $ and the variational layer $ U(\theta) $.

    \item \textbf{(6-local w/ R.)} increases locality to $k = 6$, expanding the observable space while retaining trainable rotations via $ U(\theta) $.

    \item \textbf{(6-local w/o R.)} retains the $k = 6$ observable but removes the variational layer by setting $ U(\theta) = I $, isolating the effect of the adaptive measurement.

    \item \textbf{(Only R.)} corresponds to the standard VQC in \equationautorefname{\ref{E: VQC}}, using a fixed Pauli Z gate and optimizing only the rotation parameters in $ U(\theta) $ 

\end{enumerate}

\begin{figure}[htbp]
\vskip -0.15in
\begin{center}
\centerline{\includegraphics[width=1.04\columnwidth]{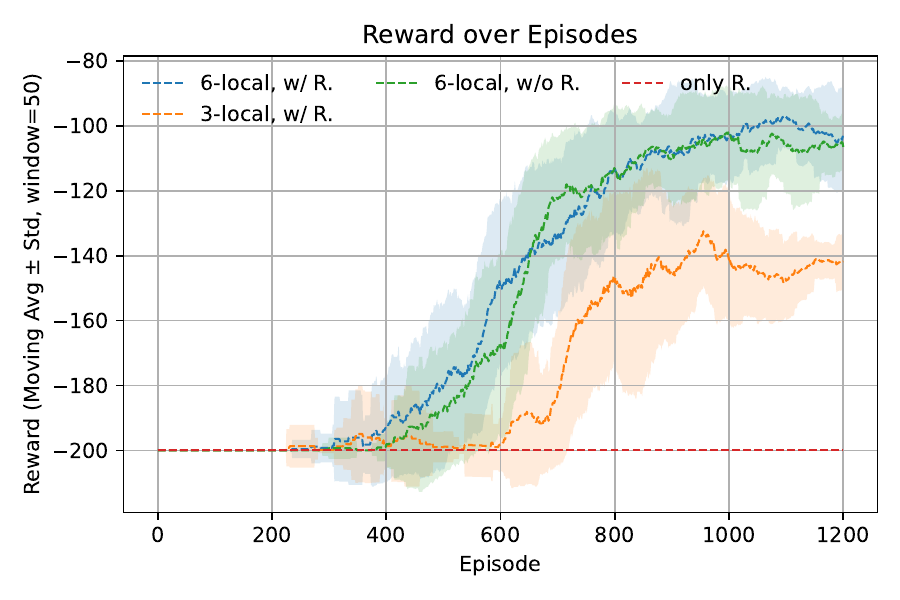}}
\caption{\textbf{[DQN] Mountain Car reward over episodes.} Increasing ANO circuit locality from 3-local to 6-local boosts learning speed and final reward. At 6-local, variants w/ R and w/o R (with and without rotation gates) converge very similarly, indicating rotation gates add negligible benefit once locality is sufficient.
}
\label{fig: dqn: mountaincar}
\end{center}
\vskip -0.3in
\end{figure}

\figureautorefname{\ref{fig: dqn: mountaincar}} shows the Mountain Car results, which indicate that the rotation-only configuration does not have enough model expressivity for this task. With ANO models, simply increasing the circuit locality from 3 to 6 yields a clear boost in performance. Moreover, at the 6-local level, adding rotation invariance provides negligible additional benefit, as both rotated and non-rotated variants converge to a similar high reward. This suggests that once the model has sufficient representational capacity (here, locality $= 6$), enforcing rotation symmetry no longer noticeably improves learning in the Mountain Car environment.

\subsection{ANO-A3C}

For the ANO-A3C setup, we integrate the ANO models directly into the A3C framework, where both the actor (policy) and critic (value) networks are realized using the quantum approximator $f_{\theta, \phi}(s, a)$ defined in \equationautorefname{\ref{E: ANO function}} and with \equationautorefname{\ref{E: ANO-A3C1}} and \equationautorefname{\ref{E: ANO-A3C2}}. This framework is benchmarked on the Cart-Pole and MiniGrid environments to examine whether ANO's increased expressivity leads to improvements in performance or training stability over traditional VQC-based agents. Across all experiments, we consider three circuit configurations: 
\begin{enumerate}
    \item \textbf{(3-local w/ R.)} 3-local ANO with variational layer, where both the variational $U$ and a 3-local adaptive observable $H$ are used.
    \item \textbf{(Only R.)} corresponds to a conventional VQC setup with $U$ and fixed Pauli Z measurement;
    \item \textbf{(Only measurement)} 3-local ANO without variational layer. That is,  $U=I$ and expressivity comes entirely from the trainable $H$.
\end{enumerate}

\subsubsection{\textbf{Cart-Pole}}

\figureautorefname{\ref{fig: a3c: cartpole}} plots the moving‐average reward ($\pm$ one standard deviation, window $= 100$) achieved by the A3C agent on the Cart-Pole task.
The 3-local w/ R. curve rises most steeply, crossing a moving‐average reward of $400$ by roughly episode $12,000$. The only Measurement variant shows slower learning, reaching only about $250$ average reward. In contrast, the only R. configuration (traditional VQC) reaches below $100$ reward. These results demonstrate that incorporating both rotation gates and non-local measurement (the 3-local w/ R. design) substantially accelerates learning and stabilizes performance in A3C on Cart-Pole.

Across both A3C and DQN Cart-Pole runs, the ANO with rotation variant outperforms rotation-only and measurement-only baselines. Additionally, A3C enables the measurement-only case to learn steadily to moderate rewards with less fluctuation, whereas it stagnates under DQN. 

\begin{figure}[htbp]
\vskip -0.15in
\begin{center}
\centerline{\includegraphics[width=1.1\columnwidth]{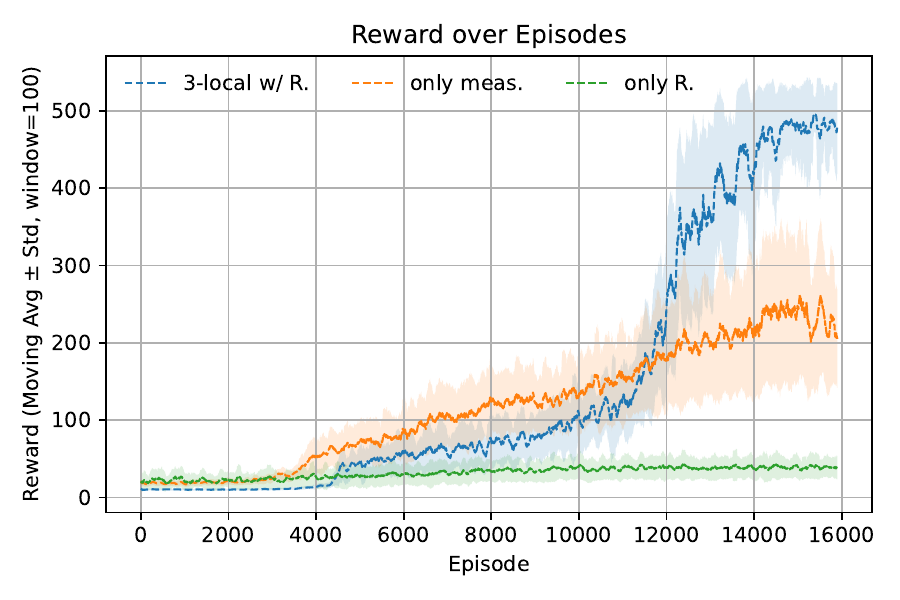}}
\caption{\textbf{[A3C] Cart-pole reward over episodes.} 
}
\label{fig: a3c: cartpole}
\end{center}
\vskip -0.2in
\end{figure}

\subsubsection{\textbf{MiniGrid}}

Two tasks are tested within the MiniGrid family: \textit{MiniGrid 8×8} and \textit{MiniGrid-SimpleCrossing S9N1}. Due to the high state dimension, a classical linear layer is employed to reduce the dimension to 4 as input features to ANO models. 

\vspace{1 mm}

\textbf{MiniGrid 8×8} is a sparse‐reward grid navigation task in which an agent must locate and reach a goal tile within an $8 \times 8$ environment, receiving +1 only upon success. In \figureautorefname{\ref{fig: a3c: minigrid8x8}}, all three circuit configurations eventually converge to about a $0.95$ success rate. Notably, ANO with rotations attains this plateau fastest with minimal variance; measurement-only follows closely, reaching around $0.95$ later; rotation-only lags behind, requiring the most episodes to reach the same level and exhibiting the highest variability.

\begin{figure}[htbp]
\vskip -0.1in
\begin{center}
\centerline{\includegraphics[width=1.1\columnwidth]{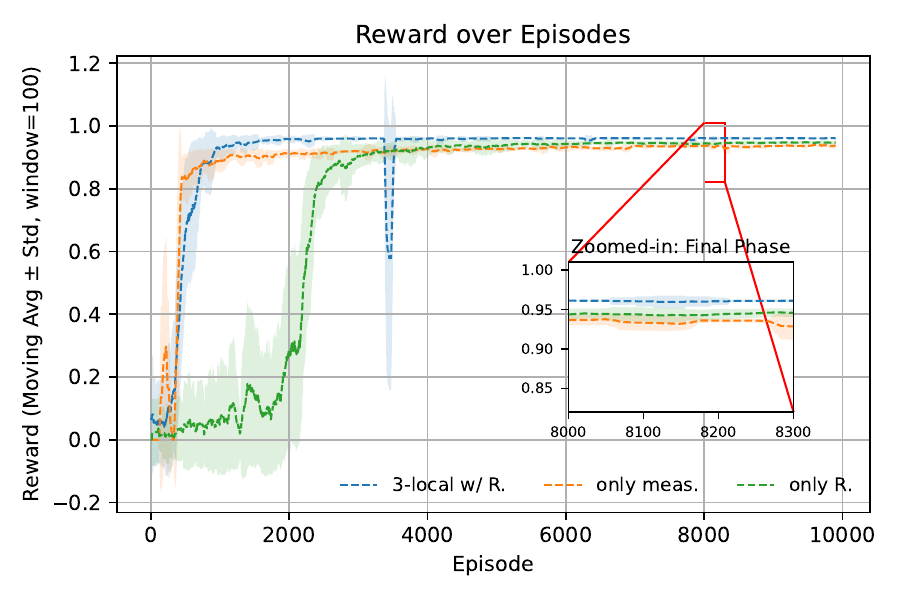}}
\caption{\textbf{[A3C] MiniGrid 8x8 reward over episodes.} 
}
\label{fig: a3c: minigrid8x8}
\end{center}
\vskip -0.1in
\end{figure}

\vspace{1 mm}

\textbf{MiniGrid SimpleCrossing S9N1} challenges the agent to traverse a narrow corridor under the same reward scheme as the $8 \times 8$ environment. \figureautorefname{\ref{fig: a3c: SimpleCrossing_S9N1}} shows that ANO with rotation climbs to above $0.8$ success by episode $6000$. The rotation-only case converges lower at around $0.4$ with greater fluctuation, while the measurement-only reaches around $0.3$, indicating the advantage of rotation gates for reliable performance in more challenging tasks.

\begin{figure}[htbp]
\vskip -0.15in
\begin{center}
\centerline{\includegraphics[width=1.1\columnwidth]{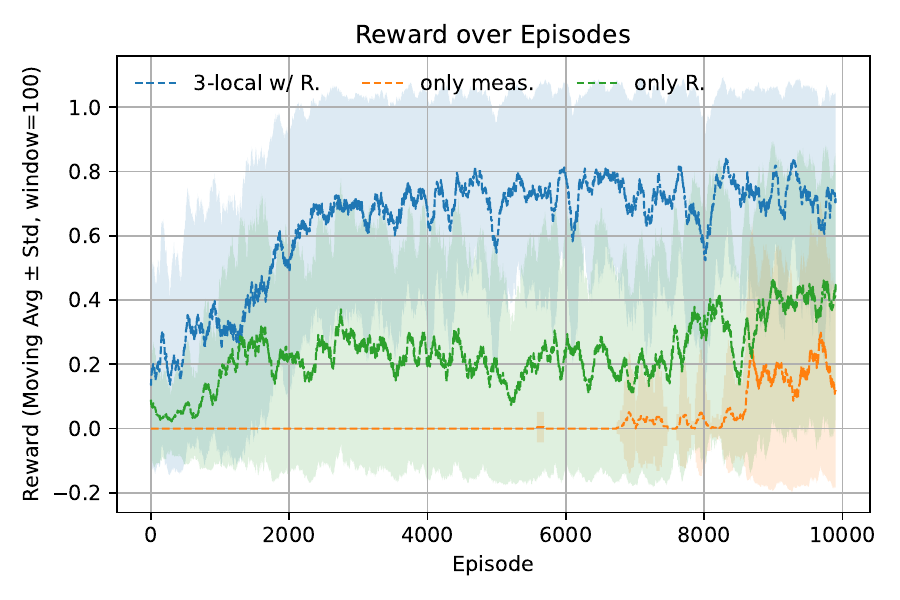}}
\caption{\textbf{[A3C] MiniGrid SimpleCrossing S9N1 reward over episodes.} 
}
\label{fig: a3c: SimpleCrossing_S9N1}
\end{center}
\vskip -0.3in
\end{figure}

\section{Conclusion}\label{sec_conclusion}
In this work, we have presented a novel integration of ANO into VQCs for QRL, embedding the ANO paradigm as the core function approximator within both DQN and A3C frameworks. By jointly optimizing circuit parameters and multi-qubit measurement operators, our ANO-VQC agents consistently outperform baseline VQCs with fixed local Pauli measurements across a suite of benchmarks, including Cart-Pole, Mountain Car, and MiniGrid variants. Experimental results demonstrate that the ANO-VQC model achieves faster convergence and higher cumulative rewards. Moreover, our ablation studies reveal that increasing measurement locality significantly expands the representational capacity of the quantum model without deepening the circuit. Notably, the measurement-only variant also learns effectively in simpler navigation tasks or with large enough locality. This highlights the flexibility afforded by adaptive observables.

Overall, our findings indicate that adaptive multi-qubit measurements can unlock latent expressive power in hybrid quantum-classical RL agents, enabling more efficient exploration of complex value landscapes on NISQ-era hardware. 

\bibliographystyle{IEEEtran}
\bibliography{references,bib/qml_examples,bib/vqc,bib/explain_qml,bib/qrl,bib/classical_rl}

\end{document}